\documentclass[12pt]{article}
\usepackage{graphicx,amsmath,amssymb,mathrsfs}
\usepackage{epsfig}

\begin{document}

\begin{center}

{\bf Axion physics in a Josephson junction environment}

\vspace{1cm}

{\bf Christian Beck}

\vspace{1cm}

{Queen Mary University of London, School of Mathematical Sciences, Mile End Road, London E1 4NS, UK}

\vspace{1cm}


\abstract{
We show that recent experiments based on Josephson junctions, SQUIDS,
and coupled Josephson qubits have a cosmological interpretation in terms
of axionic dark matter physics, in the sense that they allow for
analogue simulation of early-universe axion physics.
We discuss
new experimental setups in which SQUID-like
axionic interactions
in a resonant Josephson junction environment can be tested, similar in nature to
recent experiments
that test for quantum entanglement of two coupled Josephson qubits.
The parameter values relevant for early-universe
axion cosmology are accessible with present day's achievements in nanotechnology.}

\end{center}
\newpage

\section{Introduction}

What do advanced
Josephson junction technologies,
SQUIDs, coupled Josephson qubits
and related superconducting devices used in nanotechnology
have in common with
the problem of dark matter in the early universe? A lot more than
might seem obvious at first sight.
One of the major candidates for dark matter in the
universe is the axion.
The equation of motion of the axion
misalignment angle and that of the phase difference in a Josephson junction are identical
if the symbols in the mathematical equations are properly re-interpreted. This allows
for analogue simulation of early-universe physics
using superconducting electronic devices
such as SQUIDS and Josephson junctions.
It also allows for new experimental setups that test axionic interaction
strengths in a Josephson junction environment, similar in nature to
recent experiments
that test for quantum entanglement of two coupled Josephson qubits.
We show in this paper that the parameter values relevant for early-universe
axion cosmology are accessible with present day's achievements in nanotechnology.
Moreover, we will discuss novel types of SQUID-like interaction states by which
axionic dark matter may couple into a given resonant Josephson junction
environment. This may pave the way for novel dark matter
detectors in the future.

The research fields of superconductivity and cosmology are normally
proceeding independently of each other,
and the two scientific
communities don't know each other --- dealing apparently with very different
subject areas.
But a
look at the equations of motions of Josephson junctions,
SQUIDS, coupled Josephson qubits and similar
superconducting devices used in nanotechnology
on the one hand and axionic dark matter on the other indicates that it makes sense to think
about common approaches in both areas. The equations of motion
 are very much the same (with a suitable
re-interpretation of the currents involved) and hence it makes sense
to
develop common approaches and
to translate known results from one area (Josephson junctions) into
possible results and phenomena for the other area (axion cosmology).
This will be worked out
in this paper.

There are currently two major candidates for dark matter in the universe,
with extensive experimental searches for both of them,
WIMPS (weakly interacting massive particles)
\cite{bernabei, CDMS2, aalseth} and axions
\cite{tegmark, sikivie2, duffy, sakharov}.
In contrast to WIMPS, axions are very
light particles. In spite of their small mass,
they behave like
a very cold quantum gas \cite{sikivie2}.
There are several
experiments that search for axions in the laboratory,
using e.g.\ cavities and strong magnetic fields, which trigger the decay of axions into
two microwave photons \cite{sikivie2, duffy, cavity}.
These microwave photons are detectable if the cavity
resonates with the axion mass. Searches have been unsuccessful so far, but one needs to
scan a huge range of cavity frequencies. For this purpose,
SQUID amplifiers with very low noise levels are a very
useful technological tool to reduce the noise level
and to improve the scanning efficiency
\cite{cavity}.
These applications of low-level noise SQUID amplifiers
for axion searches are very important from an efficiency point of view
for cavity experiments, but different
from the fundamental physics analogies between axions and Josephson junctions that we want to emphasize
in this paper.
Novel ideas of experimental axion detection
have recently also been presented in \cite{ar1, ar2}, based on
cold molecules as suitable
detectors \cite{ar1} and analysis of X-rays from the sun \cite{ar2}.
Quasi-axionic particles play also an important role in
topological insulators, new materials with exotic metallic states on
their surfaces \cite{moore,wilczek,qi,bermudez}. All this
illustrates that it does make sense
to look at axion physics in a much broader
context than within
the original
model, which was
dealing with the strong CP problem in the
standard model of elementary particle physics \cite{peccei}.

The axion is described by a phase angle,
the axion misalignment angle. A Josephson junction is
also characterized by a phase, namely the phase difference of the macroscopic
wave function describing the two superconducting electrodes of the junction.
Our major motivation in this paper is the fact that
the equation of motion of the axion misalignment angle is identical to that
of the phase difference of a resistively shunted Josephson junction, with a suitable re-interpretation
of the symbols involved. As a first step, this opens up the theoretical
possibility to connect both fields, and to make analogue experiments
simulating
axion cosmology using superconducting devices in the laboratory.
Moreover, and more importantly, this novel approach also opens up the possibility to test for interaction strengths
of incoming present-day axionic dark matter in a given resonant Josephson junction environment.
The principal idea is that axions may briefly form SQUID-like interaction states
in a resonant Josephson environment, before decaying into microwave photons.
There are no cosmological constraints on these types of dark
matter interactions, since almost all of the matter
of the universe
is not in the form of Josephson junctions. The only way to either
confirm or refute such a hypothesis
is by doing laboratory experiments.
Suggested future experiments of this type can be easily performed in the laboratory and
may ultimately open up the route for novel
methods of dark matter detection based on modern nanotechnology.

This paper is organized as follows. In section 2 we point out
the equivalence of the field equations of
axions and Josephson junctions. In section 3
we show that there is quantitative agreement of the relevant
parameters. In section 4 we investigate SQUID-like interaction states
of axions and Josephson junctions. Finally, in section 5
the axionic Josephson effect is discussed.

\section{Comparing the equations of axion- and Josephson junction physics}

Let's compare the mathematics underlying both axions and Josephson junctions.
Consider an axion field $a=f_a \theta$, where $\theta$
is the axion misalignment angle and $f_a$ is
the axion coupling constant. In the early
universe, described by a Robertson Walker metric,
the equation of motion of the axion misalignment angle $\theta$
is
\begin{equation}
\ddot{\theta} +3H \dot{\theta}+ \frac{m_a^2c^4}{\hbar^2} \sin \theta = 0, \label{e1}
\end{equation}
neglecting spatial gradients.
Here $H$ is the Hubble parameter and $m_a$ denotes the axion mass. The forcing term $\sin \theta$
is due to QCD instanton effects. In a mechanical analogue, the
above equation is that of a pendulum in a constant gravitational field
with some friction determined by $H$.

When electric and magnetic fields
$\vec{E}$ and $\vec{B}$ are present as well, the axion couples as follows:
\begin{equation}
\ddot{\theta} +3H \dot{\theta}+ \frac{m_a^2c^4}{\hbar^2} \sin \theta = \frac{g_\gamma}{\pi}
\frac{1}{f_a^2} c^3 e^2 \vec{E} \vec{B} \label{1}.
\end{equation}
$g_\gamma$ is a model-dependent dimensionless coupling constant
($g_\gamma =-0.97$ for KSVZ axions,
$g_\gamma=0.36$ for DFSZ axions).
The typical parameter ranges that are allowed for dark matter axions are
\cite{sikivie2, duffy}
\begin{equation}
6 \cdot 10^{-6}eV \leq m_ac^2 \leq 2 \cdot 10^{-3} eV \label{333}
\end{equation}
and
\begin{equation}
3 \cdot 10^9 GeV \leq f_a \leq 10^{12} GeV.
\end{equation}
The
product $m_ac^2f_a$ is of the order $m_ac^2 f_a \sim 6 \cdot 10^{15} (eV)^2$.

Let's now look at
resistively shunted Josephson junctions (RSJs) \cite{josephson,tinkham}.
In the `tilted-washboard' model
the phase difference $\delta$
of the macroscopic wave function of the two superconductors satisfies
\begin{equation}
\ddot \delta +\frac{1}{RC} \dot{\delta} +\frac{2eI_c}{\hbar C}
\sin \delta = 0 \label{e2} .
\end{equation}
Here $R$ is the shunt resistance, $C$ is the capacity of the junction, and $I_c$ is the critical
current of the junction.
The frequency
\begin{equation}
\omega =\sqrt{\frac{2eI_c}{\hbar C}}
\end{equation}
is the
plasma frequency of the Josephson junction. The product
\begin{equation}
Q:= \omega RC
\end{equation}
is the quality factor of the junction.

If a bias current $I$ is applied to the junction by maintaining
a voltage difference $V$ between the two superconducting electrodes, then the equation of
motion is
\begin{equation}
\ddot \delta +\frac{1}{RC} \dot{\delta} +\frac{2eI_c}{\hbar C} \sin \delta = \frac{2e}{\hbar C} I. \label{2}
\end{equation}
Remarkably, the equations of motion of axions and of
RSJs are identical provided the following identifications are made
in eqs.~(\ref{1}) and (\ref{2}):
\begin{eqnarray}
3H&=& \frac{1}{RC} \label{3} \\
\frac{m_a^2c^4}{\hbar} &=& \frac{2eI_c}{C} \label{4} \\
\frac{g_\gamma}{\pi f_a^2} c^3 e^2 \vec{E} \vec{B} &=& \frac{2e}{\hbar C} I. \label{5}
\end{eqnarray}

An interesting consequence of this observation is the fact
that it is possible to make analogue experiments
with RSJs that
simulate axion cosmology in the laboratory. To simulate an axion
in a certain era of cosmological evolution, one builds
up an RSJ with parameters $R,C, I_c, I$ fixed by eqs.~(\ref{3})-(\ref{5}).
The left-hand side are cosmological parameters, the right-hand side is
nanotechnological engineering.
Essentially the inverse Hubble parameter $H^{-1}$ (the age of the universe) fixes the product $RC$,
the axion mass fixes the critical current $I_c$ and the axion coupling to external electromagnetic
fields $\vec{E} \vec{B}/f_a^2$ is represented by the
strength of the bias current $I$.

\section{Coincidence of axion parameters with those of superconducting
devices}

We now show that there is not only qualitative, but also quantitative
coincidence between axion and Josephson junction physics.
As worked out below, the numerical values of the parameters for
typical axionic dark matter physics
and for typical Josephson junction experiments have similar order of magnitude.
Let us start with
the experiments performed
by Koch, Van Harlingen and Clarke in \cite{koch}.
They built up four different samples
of Josephson junctions
with parameters values in the range
$R \sim 0.075-0.77 \Omega$, $C \sim 0.5-0.81 pF$, $I_c \sim 0.32-1.53 mA$.
According to eqs.~(\ref{3})-(\ref{5}), these experiments of Koch et al.
simulate
axion-like particles in a very early universe whose age is
of the order $H^{-1}=3RC \sim 10^{-13}-10^{-12}s$.
The axion mass
is in the range
$0.98 \cdot 10^{-3}-1.58\cdot 10^{-3}$ eV.
This simulated axion mass is just at the upper end of what is
expected for dark matter axions, see eq.~(\ref{333}).

The recent experiment by Steffen et al. \cite{steffen},
which tests for quantum entanglement of coupled Josephson qubits,
is classically well described by
an RSJ model where the
product $3RC$ is of the order
$H^{-1}=3RC=1.2 \cdot 10^{-6}s$ \cite{blackburn}. This experiment
thus simulates the
dynamics of weakly coupled axions
{\em after} the QCD phase transition in the early universe, which took place
during an epoch where $H^{-1}\sim 10^{-8}s$.
It corresponds
to an axion mass of $m_ac^2=\hbar \omega = 3.3 \cdot 10^{-5}eV$, much smaller
than for the Koch et al. experiment \cite{koch} but within the range expected
for dark matter axions, see eq.~(\ref{333}).

It is interesting to see that recent experiments
\cite{steffen,blackburn,penttillae,nagel,della-rocca,takahide}
simulate axions with masses in the entire range of what is interesting
from a dark matter point of view.
The experiments of Penttillae et al.
\cite{penttillae}, dealing with superconductor-insulator
phase transitions, correspond to $m_ac^2=1.32 \cdot 10^{-4}eV$.
Nagel et al. \cite{nagel} deal with negative absolute resistance effects
in Josephson junctions, these experiments formally have
$m_ac^2=2.83 \cdot 10^{-5}eV$.
Superconducting atomic contacts \cite{della-rocca} correspond to even smaller
axion masses, namely $m_a=6.7 \cdot 10^{-6}eV$,
at the lower end of what is allowed in eq.~(\ref{333}).
Two-dimensional Josephson arrays,
as built up in \cite{takahide},
correspond to arrays of coupled axions with
$m_ac^2$ in the range $6.62 \cdot 10^{-5}-1.52\cdot 10^{-4}eV$.
All these recent experiments are within
the range of axion masses that are of interest from a
dark matter point of view.
They can thus be regarded as realistic
analogue experiments simulating an axionic dark matter environment
in the early universe.

The main idea of this paper is to go a step further. If experiments
of the type mentioned above
simulate a kind of realistic axionic dynamics, could these
experiments then be used to detect current day incoming dark matter axions,
by triggering their decay in a resonant environment? This will be worked
out in the following sections.

\section{Possible interaction mechanism between
axions and Josephson junctions}


Consider a RSJ which has a plasma frequency $\omega_p=\sqrt{\frac{2eI_c}{\hbar C}}$ close to
the dark-matter axion mass $m_a$, according to eq.~(\ref{4}), and that is driven by
an external bias current $I>I_c$. Free axion quanta correspond to
small nearly-harmonic oscillations of the misalignment angle $\theta$,
in accordance with eq.~(\ref{e1}).
Suppose such an axion enters a Josephson junction that has similar parameters
as the entering axion.
Then this basically means we
have a system of two Josephson junctions, the second one represented by the entering axion.
The axion is very cold and its potential is given by a tilted washboard potential. 
Thus its phase variable may couple into a given
Josephson environment in a SQUID-like configuration. This is illustrated in Fig.~1.
\begin{figure}
\includegraphics[width=7cm]{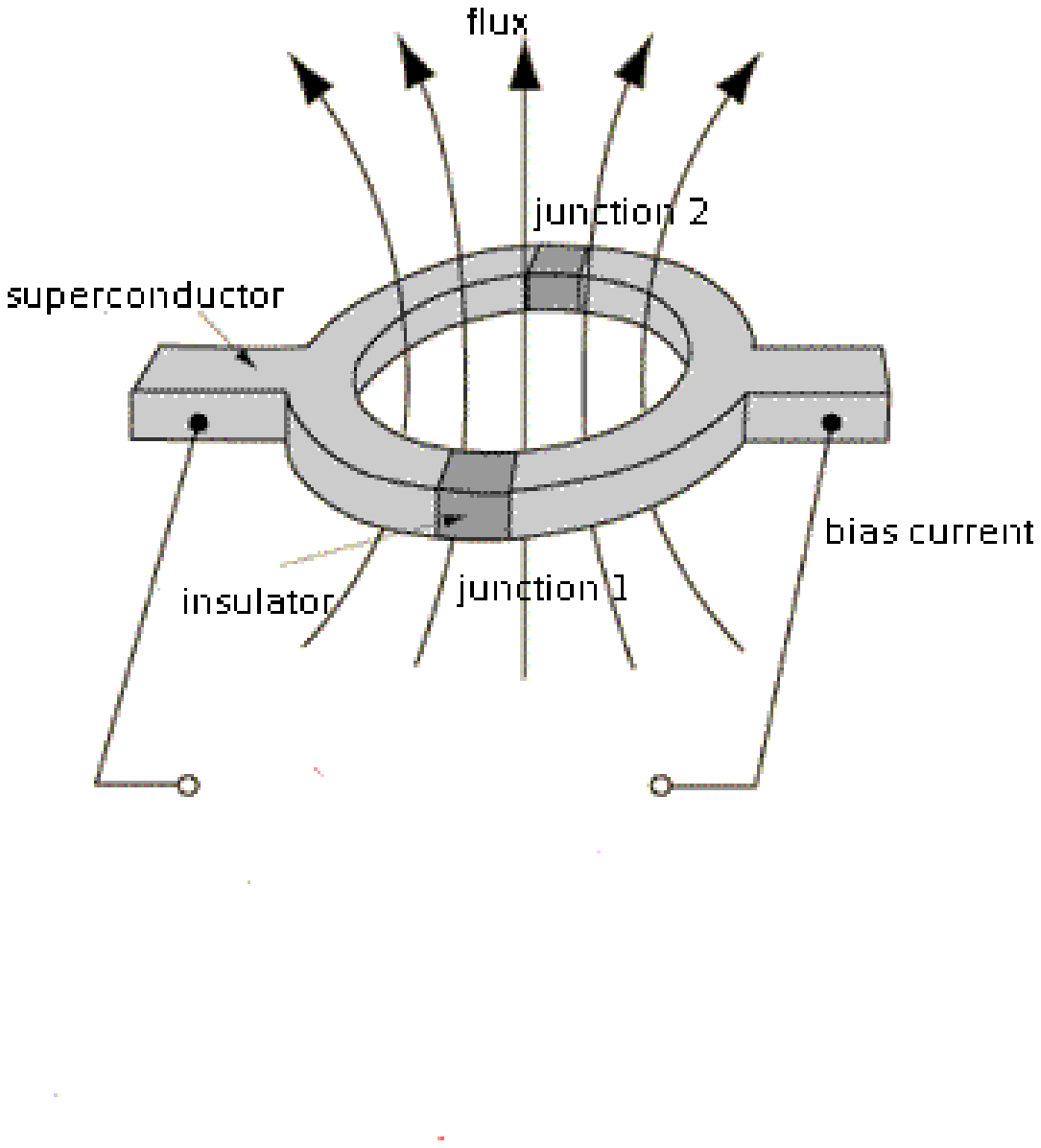}
\caption{A Josephson junction (junction 1) and an axion (formally
represented by junction 2) interacting in a SQUID
-like configuration. The supercurrent $I$ flowing
through junction 1 simulates to the synchronized axion
the formal existence of
a very large product $\vec{E}\cdot \vec{B}$ given by eq.~(\ref{5}).
Thus
the axion immediately decays into two microwave photons.}
\end{figure}

For any Josephson junction,
a given numerical value of a phase difference $\Delta \varphi$ on its own
does not have physical meaning since it is not a gauge-invariant
quantity
(see, e.g., \cite{tinkham}). Rather, what is of physical relevance
is the gauge-invariant phase difference
\begin{equation}
\delta =\Delta \varphi -\frac{2\pi}{\Phi_0} \int \vec{A} d\vec{s},
\end{equation}
where the integration path is from one electrode of the RSJ to the other ($\vec A$ is the vector potential
under consideration and $\Phi_0=\frac{h}{2e}$ is the flux quantum).
The term $\int \vec{A} d\vec{s}$ has profound consequences for the physics
of Josephson junctions and SQUIDS if magnetic fields are applied.

The gauge-invariant phase
of the Josephson-axion-SQUID of Fig.~1 must be single-valued.
Putting a closed integration path that passes
the weak links as well as the interior of the superconductor,
and using the same line of arguments as for ordinary SQUIDS
\cite{tinkham}, one obtains
\begin{equation}
\delta -\theta=2\pi \frac{\Phi}{\Phi_0} \mod 2\pi. \label{flux}
\end{equation}
Here $\Phi$ is the magnetic flux enclosed by the SQUID.
If the flux $\Phi$
is close to zero or given by integer multiples of $\Phi_0$,
then the above condition
simply implies
\begin{equation}
\delta = \theta ,
\end{equation}
i.e.\ the two phases synchronize.

Classically,
the coupling between two Josephson junctions with
shunt resistance $R$, capacity $C$, and inductivity $L$ is
described by the following coupled differential equations
\cite{blackburn}:
\begin{eqnarray}
\ddot \delta +\frac{1}{RC} \dot \delta +\omega^2 \sin \delta &=&\gamma_x (\ddot \theta -\ddot \delta) +\frac{1}{CL}
(\delta +2\pi M_1) \nonumber \\
\ddot \theta +\frac{1}{RC} \dot \theta +\omega^2 \sin \theta &=&\gamma_x (\ddot \delta -\ddot \theta) +\frac{1}{CL}
(\theta +2\pi M_2) \nonumber \\
\, & \, & \, \label{coupled}
\end{eqnarray}
 Here $\delta$ is the phase difference of the first junction, $\theta$ that of the
second junction.
$M_i=\Phi_i/\Phi_0$ is the normalized flux
enclosed by junction $i$ ($i=1,2$), and $\gamma_x=C_x/C$ is a
small dimensionless coupling constant, assuming both
junctions are capacitively coupled by a capacity $C_x$.
For example, in the experiments of Steffen et al.
dealing with coupled Josephson qubits \cite{steffen} one has $\gamma_x = 2.3 \cdot 10^{-3}$.
Typically, the
damping term proportional to $\dot \delta$ and $\dot \theta$ is neglected in
the theoretical treatment of these types of experiments \cite{blackburn}.
The above classical equations of motion describe
quite well the experimentally observed
phenomena \cite{steffen, blackburn}.

In our case
the phase $\delta$ describes
an ordinary Josephson junction and the phase $\theta$
an axion that passes through this Josephson junction.
A very simple coupling scheme is given by
\begin{eqnarray}
\ddot \delta +\frac{1}{RC} \dot \delta +\omega^2 \sin \delta &=&\gamma_x (\ddot \theta -\ddot \delta)
\nonumber \\
\ddot \theta +3H \dot \theta +\frac{m_a^2c^2}{\hbar^2} \sin \theta &=&\gamma_x (\ddot \delta -\ddot \theta) .
\label{coupled2}
\end{eqnarray}
This corresponds to the case $L\to \infty$ in eq.~(\ref{coupled}).

If the axion mass is at resonance with the Josephson
plasma frequency, $m_ac^2=\hbar \omega$,
then synchronization effects of the phases $\delta$ and $\theta$ will occur
if $\gamma_x$ is not too small,
just as they occur for coupled Josephson qubits \cite{steffen,
blackburn}.
If the axion couples to fluxes similar as in eq.~(\ref{coupled}), then
this is described by a small additional self-interaction potential
given by
$V(a)=-\frac{1}{CL}(\frac{1}{2}a^2+2\pi M_2 f_a a)$.
Quantum mechanically, one could even
speculate on the formation of
entangled states between axions and Josephson qubits.

Given the quantitative agreement between
the parameters of axion physics and Josephson junction physics outlined
in section 3, one might hope that for axions the coupling
$\gamma_x$ is again of similar order of magnitude
as in current nanotechnological experiments, provided the
Josephson plasma frequency is close to the axion mass. This can be experimentally tested.

There are no astronomical
constraints on the size of $\gamma_x$
 since almost all of the matter in the universe is not
in the form of Josephson junctions.
Axionic dark matter may look completely `dark' in the universe as a whole but not at all
`dark' in special superconducting devices designed by mankind,
due to SQUID-like interactions.
The only way to constrain $\gamma_x$
is to scan a range of plasma frequencies and look for the
possible occurence or non-occurence
of universal resonance effects, produced by axions of the dark matter halo
that hit terrestrial Josephson junction experiments.
The intensity of this effect might display small yearly modulations, just similar as in
the DAMA/LIBRA experiments \cite{bernabei}.
 What corresponds to tuning the cavity
frequency in the experiments \cite{cavity} would correspond to tuning
the plasma frequency $\omega$ in these new types of
nanotechnological dark matter experiments.

\section{Axionic Josephson effect}

As an application of our theoretical treatment in the previous section,
let us now discuss the analogue of the Josephson effect for axions,
similar in spirit
to what was experimentally observed for BEC in \cite{levy}.
A Josephson junction biased with voltage $V$ generates Josephson radiation with frequency
\begin{equation}
\hbar \omega_J =2eV.
\end{equation}
For such a biased junction the phase $\delta$ grows linearly in time, i.e.
\begin{equation}
\delta (t) = \delta
 (0) +\frac{2eV}{\hbar} t.
\end{equation}
The relation between bias current $I$ and applied voltage $V$ is
\begin{equation}
V= R \sqrt{I^2-I_c^2} \approx RI \mbox{$\;\;\;$ for $I>>I_c$}.
\end{equation}
Josephson oscillations set in if
\begin{equation}
I>I_c, \label{ic}
\end{equation}
i.e. the bias current $I$ must be larger than the critical current $I_c$
of the junction. In the mechanical analogue, the pendulum rotates
with large kinetic energy.

According to eq.~(\ref{1}) and (\ref{2}), the axion
misalignment angle $\theta$ will also start to increase linearly in time if it is being
forced by very strong products of $\vec{E}$ and $ \vec{B}$ fields. So from
a formal mathematical point of view, an
axionic Josephson effect is possible. The axionic Josephson frequency is given by
\begin{equation}
\hbar \omega_J =2eV \approx 2e RI = \frac{g_\gamma}{\pi} \frac{1}{f_a^2} c^3 \vec{E} \cdot
\vec{B}, \label{11}
\end{equation}
where in the last step eq.~(\ref{5}) was used.
Condition (\ref{ic}) translates to
\begin{equation}
e^2 \hbar^2 \vec{E} \vec{B} > f_a^2 m_a^2 c = \Lambda^4 c \label{ic2}
\end{equation}
QCD-inspired models of axions fix $\Lambda$ to be about $78$ MeV \cite{tegmark}.
Strong magnetic fields in the laboratory correspond to about $10$
Tesla, and strong electric fields
to about $10^9V/m$. This gives $\vec{E} \cdot \vec{B} \approx 10^{10}VT/m$. One can easily check that under normal laboratory conditions one cannot
produce stationary $\vec{E}$ and $\vec{B}$ fields of sufficient strength to satisfy eq.~(\ref{ic2}).


However, there is another interesting possibility how we can briefly
induce axionic Josephson oscillations in the lab. This is based on
the SQUID-like interaction mechanism discussed in section 4.
Remember that the phase of a SQUID formed
out of a Josephson junction and a passing axion should be gauge-invariant,
provided the axion mass is at resonance with the plasma frequency.
This led us to derive eq.~(\ref{flux}), meaning that the phase $\delta$
of the Josephson junction and the phase $\theta$ of the axion synchronize.
For a biased RSJ performing Josephson oscillations of
frequency $\omega_J$, synchronization means that
\begin{equation}
\delta (t) = \delta (0)+ \omega_J t
\end{equation}
induces
\begin{equation}
\theta (t) = \theta (0)+ \omega_J t
\end{equation}
for the axion.
According to eq.~(\ref{5}), this means that the axion
formally sees a huge product field $\vec{E} \cdot \vec{B}$.
The huge (virtual) magnetic field will make it immediately decay
into two microwave photons. The microwave photons produced by axion
decay have the frequency of the Josephson radiation and produce
distortions in the $I-V$ curve.
They can be potentially measured in
form of Shapiro steps (Shapiro steps are well-known step-like structures in
the $I-V$ curves of irradiated Josephson junctions \cite{shapiro}).
Our theoretical idea thus opens up the possibility to develop
new detectors for
axionic dark matter\footnote{
A strong resonance of unknown origin, observed in the experiments of
Koch et al.\cite{koch} at 368 GHz for the 4th junction
of their experimental series would point to an axion
mass of $1.52 \cdot 10^{-3}$ eV if interpreted in this way.}.

\section{Conclusion}

Let us conclude. In this
paper we  have discussed the possibility
that axions could interact with Josephson junctions
by briefly forming SQUID-like interaction states.
Josephson junctions, SQUIDS,
and spatially extended arrays of these superconducting devices can nowadays be built
for a wide range of different parameters $R,C,I_c$, and it is very easy to tune
the bias current $I$ to any value of interest.
It is also very easy to adjust the plasma frequency of a Josephson
junction to any value of interest.
It is thus possible to
build up a suitable resonant environment that could help to detect
incoming dark matter axions.

There is the prospect of
developing new generations of detectors for dark matter axions that
search for possible resonance effects and phase synchronization
if the Josephson plasma frequency is close to the axion mass.
In this way the size of the coupling $\gamma_x$ between
axions and a given Josephson environment can be
experimentally constrained.
As shown in section 3, the
relevant dark matter mass parameter range is accessible by modern technological
developments in nanotechnology.
An obvious advantage of these types of experiments is that
the formal existence of extremely large products of electric and magnetic field strengths
$\vec{E} \vec{B}$ acting on the axion can be simulated by a very simple
experimental setup, an easily tunable bias current $I$, assuming that
some axions hitting the Josephson junction will synchronize their phase
due to a SQUID-like interaction.
This effect may be systematically tested in future experiments.

\end{document}